\documentclass[aps,prl,twocolumn,superscriptaddress]{revtex4}

\usepackage{graphicx}
\usepackage{amsmath}
\usepackage{amssymb}
\usepackage{color}
\usepackage{braket}

\begin{document}

\title{Universal Feature in Optical Control of a $p$-wave Feshbach Resonance}

\author{Peng Peng}
\thanks{They contribute equally to this work. }
\affiliation{State Key Laboratory of
Quantum Optics and Quantum Optics Devices, Institute of
Opto-Electronics, Shanxi University, Taiyuan 030006, P.R.China }
\affiliation{Collaborative Innovation Center of Extreme
Optics, Shanxi University, Taiyuan 030006, P.R.China}
\author{Ren Zhang}
\thanks{They contribute equally to this work. }
\affiliation{Department of Applied Physics, Xi'an Jiaotong University, Shanxi, 710049, China}
\affiliation{Institute for Advanced Study, Tsinghua University, Beijing, 100084, China}
\author{Lianghui Huang}
\thanks{Correspondence may be addressed to Lianghui Huang (huanglh06@126.com).}
\affiliation{State Key Laboratory of Quantum Optics and Quantum
Optics Devices, Institute of Opto-Electronics, Shanxi University,
Taiyuan 030006, P.R.China } \affiliation{Collaborative Innovation
Center of Extreme Optics, Shanxi University, Taiyuan 030006,
P.R.China}
\author{Donghao Li}
\affiliation{State Key Laboratory of
Quantum Optics and Quantum Optics Devices, Institute of
Opto-Electronics, Shanxi University, Taiyuan 030006, P.R.China }
\affiliation{Collaborative Innovation Center of Extreme
Optics, Shanxi University, Taiyuan 030006, P.R.China}
\author{Zengming Meng}
\affiliation{State Key Laboratory of
Quantum Optics and Quantum Optics Devices, Institute of
Opto-Electronics, Shanxi University, Taiyuan 030006, P.R.China }
\affiliation{Collaborative Innovation Center of Extreme
Optics, Shanxi University, Taiyuan 030006, P.R.China}
\author{Pengjun Wang}
\affiliation{State Key Laboratory of
Quantum Optics and Quantum Optics Devices, Institute of
Opto-Electronics, Shanxi University, Taiyuan 030006, P.R.China }
\affiliation{Collaborative Innovation Center of Extreme
Optics, Shanxi University, Taiyuan 030006, P.R.China}
\author{Hui Zhai}
\affiliation{Institute for Advanced Study, Tsinghua University, Beijing, 100084, China}
\affiliation{Collaborative Innovation Center of Quantum Matter, Beijing, 100084, China}
\author{Peng Zhang}
\affiliation{Department of Physics, Renmin University of China, Beijing, 100872,
China}
\affiliation{Beijing Computational Science Research Center, Beijing, 100084, China}
\author{Jing Zhang}
\thanks{Correspondence may be addressed to Jing Zhang (jzhang74@yahoo.com,
jzhang74@sxu.edu.cn).}
\affiliation{State Key Laboratory of
Quantum Optics and Quantum Optics Devices, Institute of
Opto-Electronics, Shanxi University, Taiyuan 030006, P.R.China }
\affiliation{Synergetic Innovation Center of Quantum
Information and Quantum Physics, University of Science and
Technology of China, Hefei, Anhui 230026, P. R. China}

\date{\today}
\begin{abstract}
In this Letter we report the experimental results on optical control
of a $p$-wave Feshbach resonance, by utilizing a laser driven
bound-to-bound transition to shift the energy of closed channel
molecule. The magnetic field location for $p$-wave resonance as a
function of laser detuning can be captured by a simple formula with
essentially one parameter, which describes how sensitive the
resonance depends on the laser detuning. The key result of this work
is to demonstrate, both experimentally and theoretically, that the
ratio between this parameter for $m=0$ resonance and that for $m=\pm
1$ resonance, to large extent, is universal. We also show that this
optical control can create intriguing situations where interesting
few- and many-body physics can occurs, such as a $p$-wave resonance
overlapping with an $s$-wave resonance or three $p$-wave resonances
being degenerate.

\end{abstract}
\maketitle

The capability of controlling the interaction strength between atoms
has led to tremendous progresses in the field of ultracold atomic
gases. Magnetic-field-induced Feshbach resonance is one of such
powerful tools and has been widely used in studying strongly
correlated degenerate atomic gases  \cite{Chin2010}. Another
technique for tuning interatomic interactions is the optical
Feshbach resonance, in which a pair of atoms in the scattering
states are coupled to an excited molecular state by a near
photoassociation resonance laser field  \cite{Fedichev1996}. The
optical Feshbach resonance offers a more flexible spatial and
temporal control of interaction, since the laser intensity can vary
on short length and time scales \cite{Takahashi2010,Zhai2011}.
However, it also suffers from rapid losses of atoms due to the
light-induced inelastic collisions between atoms.

Recently, an alternative method of optical control has been
implemented to avoid the problem of atom losses, in which the
optical control is combined with the magnetic Feshbach resonance
\cite{Zhang2009,Bauer2009,Bauer2009-PRA,Grimm2005,Wu2012,Fu2013,Chin2015,Thomas2016,ZP2017,Hu2014}.
The key idea is that, instead of coupling atoms in scattering states
to a bound state, the laser induces a bound-to-bound transition
between the closed channel molecule responsible for a magnetic
Feshbach resonance and an excited molecular state. In this way, the
laser can shift the energy of closed channel molecule, and thus
moves the location of the magnetic Feshbach resonance. This method
has been recently demonstrated in both ultracold Bose
\cite{Bauer2009,Chin2015,Grimm2005} and Fermi gases
\cite{Fu2013,Thomas2016}. It has been shown that the atom loss rate
can be reduced by an order of magnitude, while the advantage of high
resolution of spatial and temporal control is still maintained. To
distinguish this method from the conventional optical Feshbach
resonance, we shall refer to it as the optical control of a magnetic
Feshbach resonance.

$p$-wave interaction plays a crucial role in many quantum many-body
systems \cite{read,jason,yu,yoshida,luciuk,Qi,cui17}, especially in
the realization of  topological superfluids
\cite{kitaev,topo2,topo3}. Thus, the intensively experimental
efforts have been made on $p$-wave Feshbach resonance  in the last
decade
\cite{pwaveFR,pwaveFR2,pwavejzhang,esslinger2005,julienne2005,jin2007,vale2008,mukai2008,mukai2013}.
In this Letter, we apply the method of optical control, for the
first time,  to a high partial wave magnetic Feshbach resonance. We
will highlight the universal features in such an optical control. In
addition, we will show that interesting situations such as
degenerate $p$-wave and $s$- and $p$-wave overlapping resonances can
indeed be created.

\begin{figure}[tp]
\includegraphics[width=3.2 in]
{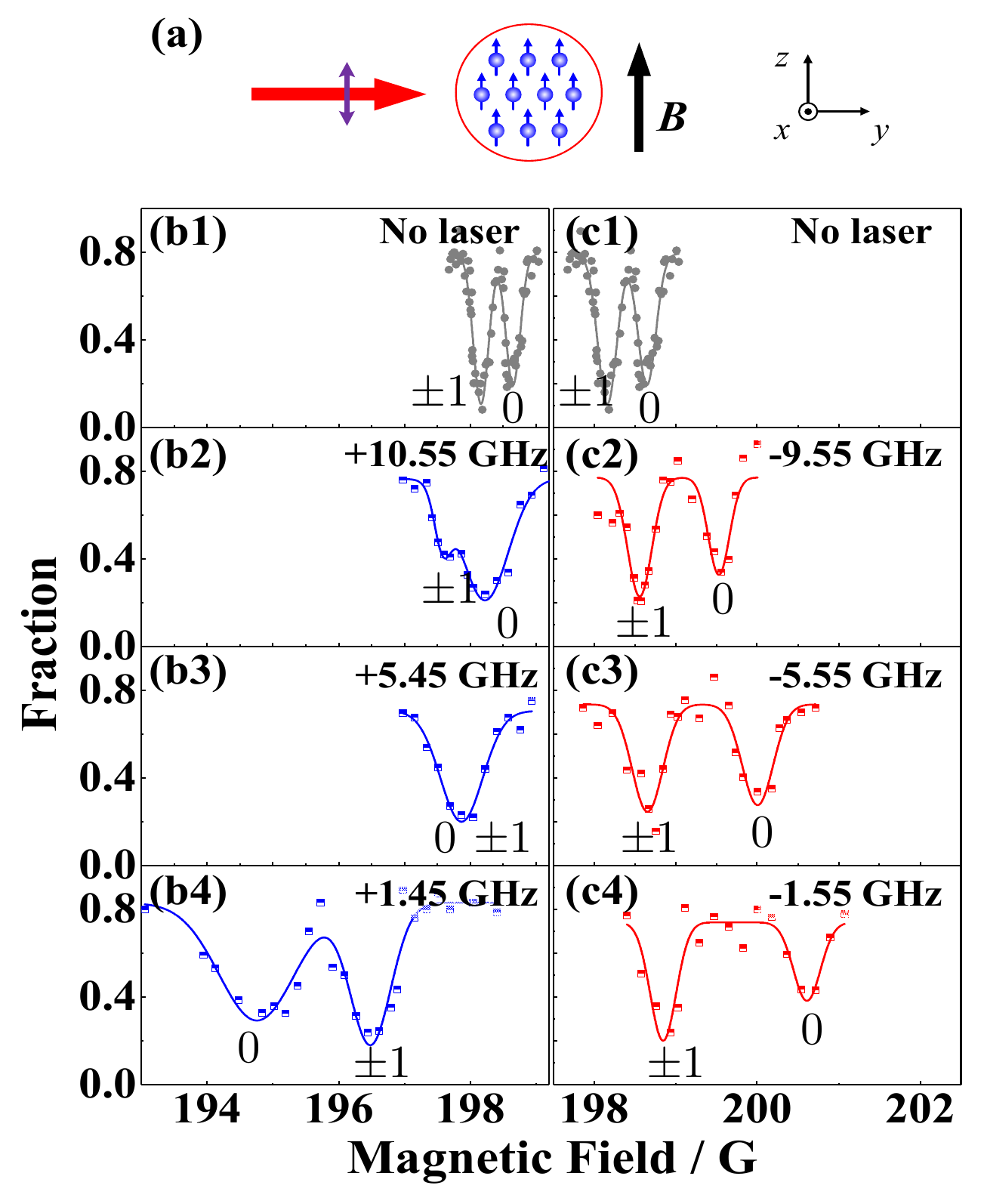} \caption{(Color online) \textbf{The $p$-wave
Feshbach resonance manipulated by a laser field with polarization
parallel to  the external magnetic field.} \textbf{(a)} Schematic
diagram of the laser beam and the external magnetic field.
\textbf{(b)} and \textbf{(c)} Atom loss measurements of the $p$-wave
Feshbach resonance of $|9/2,-7/2\rangle\otimes|9/2,-7/2\rangle$
(located at $198.3$ G for $m=\pm 1$ and $198.8$ G for $m=0$ without
laser field) as the function of the magnetic field for the different
blue (b) and red laser (c) detuning. The laser field drives a
bound-to-bound transition around $\omega_{\text{eg}}\simeq388104.6$
GHz.} \label{optical}
\end{figure}

\textit{Experimental Setup.} Our experiment is performed with
${}^{40}$K Fermi gas of $F=9/2$ manifold with atom number $N=2\times
10^{6}$ and at temperature $T/T_{F}\approx 0.3 $. Details of our
setup and state preparations can be found in Ref. \cite{Fu2013,sup}.
We start with the $p$-wave resonance for two atoms in
$|9/2,-7/2\rangle\otimes|9/2,-7/2\rangle$ with a magnetic field
along $\hat{z}$ direction. Because of the magnetic dipolar
interaction, the $m=0$ resonance occurs at a slightly higher field
of $198.8$ G and $m=\pm 1$ resonance at a slightly lower field of
$198.3$ G, as shown in Fig. \ref{optical}(b1) and (c1). We first
consider the situation that a laser with linear polarization along
$\hat{z}$ is applied to the sample, as shown in Fig.
\ref{optical}(a). Both the rotational and the time reversal symmetry
are still preserved, under the condition that the photon recoil
energy is sufficiently weak compared to the detuning and can be
safely ignored. Hence, the $m=\pm 1$ resonances remain degenerate
even in the presence of the optical control laser.


\textit{Optical Shift of Resonance Position.} When the laser is red
detuned to the bound-to-bound transition, the energy of the closed
channel bound state is effectively pushed down due to the coupling
to the excited molecular state. Consequently, it requires larger
Zeeman energy to bring the bound state to threshold, and the
Feshbach resonance moves toward high magnetic fields. When the laser
detuning becomes smaller, the bound state energy experiences
stronger level repulsion. As a result, the shift of the resonance
position becomes larger. As shown in Fig. \ref{optical}(c), we find
that the position of $m=0$ resonance moves much faster than those of
the $m=\pm 1$ resonances as the detuning decreases. For instance,
when the red detuning is $\sim 1.55$ GHz, as shown in Fig.
\ref{optical}(c4), the $m=\pm 1$ resonance is only shifted to
$198.8$ G, and the $m=0$ resonance is shifted to $201.6$ G.

\begin{figure*}
\includegraphics[width=15cm]{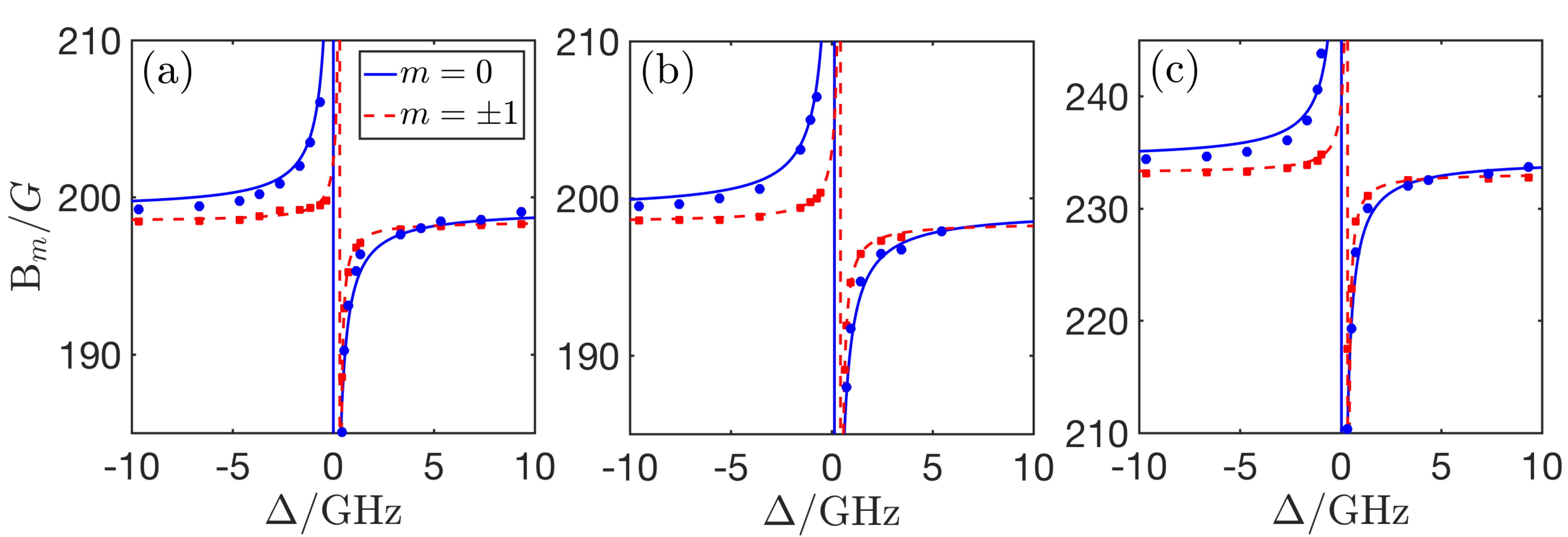}
\caption{(Color online) \textbf{The position of the shifted Feshbach
resonance as a function of the laser detuning.} (a) and (b) $p$-wave
Feshbach resonance of two atoms in
$|9/2,-7/2\rangle\otimes|9/2,-7/2\rangle$ state \cite{pwaveFR} at
about $198$ G with bound-to-bound transition frequency at
$\omega_{\text{eg}}\simeq388.105$ THz (a) and
$\omega_{eg}\simeq388.31$ THz (b); (c) $p$-wave Feshbach resonance
of two atoms in $|9/2,-5/2\rangle\otimes|9/2,-5/2\rangle$ state
\cite{ludewig} at about 232 G with bound-to-bound transition
frequency at $\omega_{eg}\simeq388.105$ THz . The curves are
obtained by fitting experimental data by Eq. (\ref{resonancepot}). }
\label{universal}
\end{figure*}

When the laser is blue detuned, the energy of the closed channel
molecule is effectively pushed to the higher energy, and it
therefore requires less Zeeman energy to bring the molecule to
threshold, and consequently the Feshbach resonances move toward low
magnetic fields. Similarly, the resonance of $m=0$ moves faster.
Hence, for small detuning, the $m=0$ resonance locates at a lower
field than $m =±1$ resonance, as shown in Fig. \ref{optical}(b4).
Nevertheless, when the detuning becomes larger, it will eventually
recover the situation in absence of the laser field, that is, the $m
= 0$ resonance locates at a higher field than the $m =\pm 1$
resonance, as shown in Fig. \ref{optical}(b2). Hence, at an
intermediate fine-tuned detuning, the $m=0$ resonance will
accidentally overlap with the $m=\pm1$ resonance, and it creates
another interesting situation that all three resonances appear as a
single resonance. This is indeed observed as shown in Fig.
\ref{optical}(b3). In another word, this happens when the difference
in dipolar interaction energy between molecules is canceled by the
difference in the molecular ac Stark effect. In this accidental
situation, it can be viewed as the SU(2) rotational symmetry is
restored

\textit{Experimental Observation of the Universal Feature.} To
clearly visualize how the resonance position is shifted by the
laser, in Fig. \ref{universal} we plot the magnetic field location
as a function of the laser detuning. In Fig. \ref{universal}(a) we
consider the $p$-wave resonance for spinless ${}^{40}$K in
$|9/2,-7/2\rangle$ state \cite{pwaveFR} and the bound-to-bound
transition at $\omega_\text{eg}\simeq 388.105$ THz, and in Fig.
\ref{universal}(b) we consider the same $p$-wave resonance but a
different bound-to-bound transition at $\omega_\text{eg}\simeq
388.31$ THz. In Fig. \ref{universal}(c) we consider a different
$p$-wave resonance for spinless ${}^{40}$K in $|9/2,-5/2\rangle$
state \cite{ludewig} with the bound-to-bound transition at a similar
frequency as Fig. \ref{universal}(a).

In Fig. \ref{universal} we also show that all these cases can be well captured by a simple formula as
\begin{eqnarray}
\label{resonancepot}
\mu(B_{m}-B_{m0})=-{\rm Re}\left[\frac{I(m)}{\Delta-i\gamma/2}\right],\label{bm}
\end{eqnarray}
where $B_m$ and $B_{m0}$ are the magnetic field position for
resonances in presence and in absence of the laser field,
respectively, $\mu$ is the magnetic moment difference between the
closed and open channels, $\Delta$ is the laser detuning from the
excited molecular states, and $\gamma$ is the spontaneous emission
rate of excited molecular states. $I(m)$ represents the laser
induced coupling between closed-channel molecule and the excited
molecular state. The R.H.S of Eq. (\ref{bm}) is nothing but the
laser-induced energy shift of the closed-channel bound state. This
formula can be derived from microscopic coupled channel model
\cite{peng,Paul2006}.

By fitting the data shown in Fig. \ref{universal} with
Eq.(\ref{resonancepot}), we find $I(0)/I(\pm1)\approx2.1$, $1.9$ and
$1.9$ for Fig. \ref{universal}(a-c), respectively. This strongly
indicates that \textit{$I(0)/I(\pm 1)$ is a universal number}. Here
\textit{universal} means this ratio is not sensitive to either the
choice of closed molecule, that is, which $p$-wave resonance to
start with (e.g. Fig. \ref{universal}(a) and (c)), or the choice of
the excited molecular state, that is, which bound-to-bound
transition to couple to (e.g. Fig. \ref{universal}(a) and (b)).

\textit{Theoretical Explanation of the Universal Feature.} Here we
offer a theoretical explanation why $I(0)/I(\pm 1)$ is indeed
universal. To start with, let us state the necessary quantum numbers
to describe a diatomic molecular state in the center-of-mass frame.
${\bf r}_n$ denotes the displacement between two nucleus, and ${\bf
r}_{\text{e}_1}$ and ${\bf r}_{\text{e}_2}$ are the displacements
between two electrons and the center-of-mass. The necessary quantum
numbers includes: (i) the total angular momentum $l$ and its
$\hat{z}$-component $m$ (Here $\hat{x}$, $\hat{y}$ and $\hat{z}$
label directions in the laboratory frame); (ii) the projection of
${\bf L}_{{\rm e}_1}+{\bf L}_{{\rm e}_2}$ along the direction of
$\hat{{\bf r}}_n$, denoted by $\lambda$; (iii) $n_\text{n}$ denoting
the vibration between two nucleus, and a set of quantum numbers
$\{n_\text{e}\}$ describing the vibration of two electrons; and (iv)
the quantum numbers describing the electron and nuclear spin degree
of freedom.

For the problem considered here, it is quite reasonable to make
following assumptions: (i) The energy splitting between different
spin states are much smaller comparing to the laser detuning, such
that we can ignore the spin-orbit coupling and the hyperfine
coupling, and therefore we will not explicitly include the electron
and nuclear spin degree of freedoms. (ii) The energy splitting
between states with different quantum number $l$ are also considered
to be small comparing to the laser detuning, and therefore, we treat
them as ``degenerate" states in the laser coupling. (iii) The radial
wave functions of the excited molecular states are not sensitive to
the quantum number $l$ and $m$. (ii) and (iii) are essentially based
on the consideration that the molecules involved in this process are
deeply bound such that their wave function are largely reside in the
centrifugal barrier.

Moreover due to the rotational symmetry along $\hat{z}$, $m$ is a
good quantum number between initial and final states. With (i) and
(ii), the laser coupling between closed and excited molecular states
is proportional to
\begin{equation}
I(m)\propto\sum_{l^\text{f}}
|\langle l^\text{f}, m, \lambda^\text{f},\{n^\text{f}_\text{e}\}, n_\text{n}^\text{f}|\hat{T}_{0}|l^\text{i}, m, \lambda^\text{i}, \{n^\text{i}_\text{e}\}, n_\text{n}^\text{i} \rangle|^2,\label{WE}
\end{equation}
where $\hat{T}_0$ denotes the $\hat{z}$-component of ${\bf
r}_{\text{e}_1}+{\bf r}_{\text{e}_2}$, and f and i in the upper
superscript label the quantum numbers for the initial and final
state quantum numbers, respectively. Since we consider a $p$-wave
resonance, the closed channel molecule should be a $p$-wave one,
that is, $l^\text{i}=1$; and for two alkali atoms in the electronic
ground state ($\Sigma$-orbital), $\lambda^\text{i}=0$. In the
expression for $I(m)$, different choice of $n_\text{n}^\text{i}$ and
$\{n^\text{i}_\text{e}\}$ corresponds to different closed channel
molecules, and thus, different $p$-wave resonance; and different
choice of $n_\text{n}^\text{f}$ and $\{n^\text{f}_\text{e}\}$
corresponds to different excited state molecules, and thus,
different bound-to-bound transition frequency.

The key theoretical result is to show that $I(m)$ can be factorized into
\begin{equation}
\label{I-equ}
I(m)=g\left(m,\lambda^\text{f}\right)\times h\left(\{n^\text{f}_\text{e}\}, n_\text{n}^\text{f}, \{n^\text{i}_\text{e}\}, n_\text{n}^\text{i},\lambda^{\rm f}\right),
\end{equation}
where $g$ and $h$ are two functions. This result follows from (iii)
and the use of the Born adiabatic approximation \cite{sup}. Thus, we
can see that $I(0)/I(\pm 1)$ only depends on $\lambda^\text{f}$ and
\begin{eqnarray}
\label{ratio}
\frac{I(0)}{I(\pm 1)}=\left\{ \begin{array}{c}
3 {\rm \ for\ \lambda^\text{f}=0},\\
\\
1/2 {\rm \ for\ \lambda^\text{f}=\pm1}.
\end{array}\right.
\end{eqnarray}
$I(0)/I(\pm 1)$ is independent of $n_\text{n}^\text{f}$,
$\{n^\text{f}_\text{e}\}$, $n_\text{n}^\text{i}$ and
$\{n^\text{i}_\text{e}\}$, that is to say, is independent on the
choice of $p$-wave resonance, and up to these two different values,
is independent of the choice of the bound-to-bound transition.

This result provides a qualitative explanation of the experimental
observations. Assuming the three cases shown in Fig. \ref{universal}
all come from excited molecular states with $\lambda^\text{f}=0$, it
is consistent with that fact that the $m=0$ resonance always moves
faster than $m=\pm 1$ resonance, and the ratio $I(0)/I(\pm 1)$ is
nearly a constant. The quantitative difference between our
theoretical and experimental results is likely due to the
assumptions (i)-(iii) are not perfectly obeyed in practices.

\begin{figure}[t]
\begin{centering}
\includegraphics[width=3in]{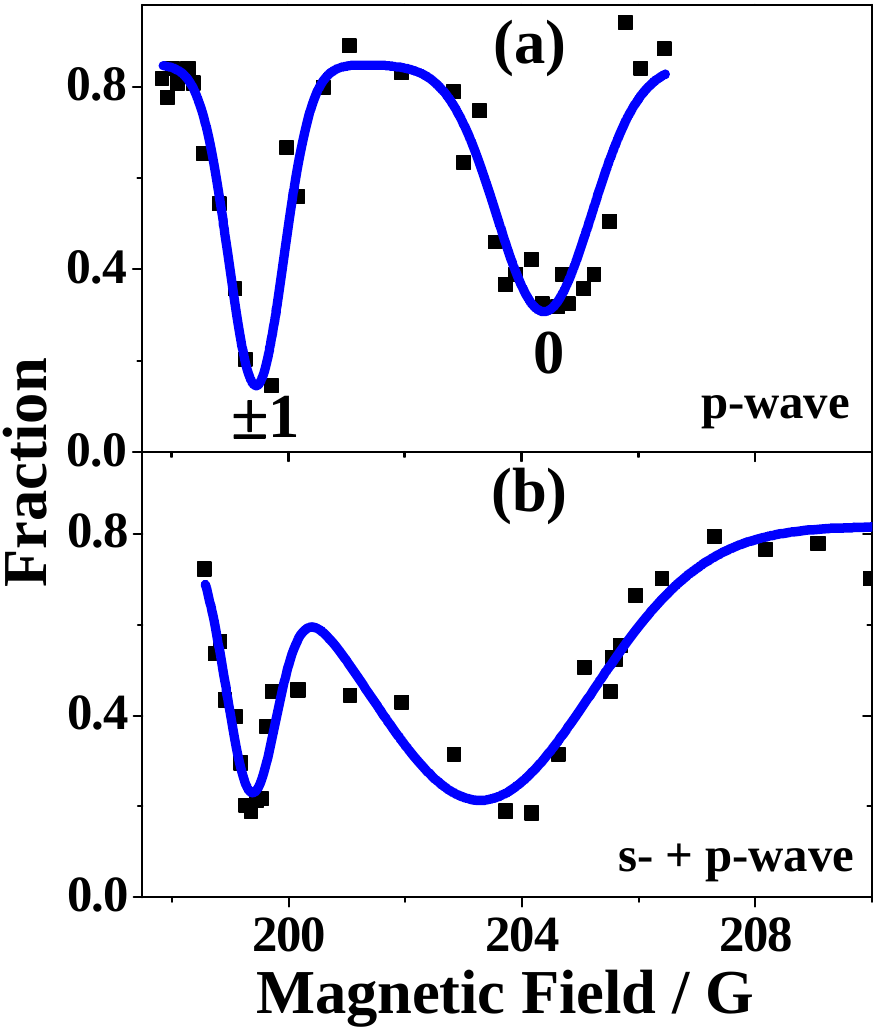}
\par\end{centering}

\caption{(Color online) \textbf{The p-wave Feshbach resonance
overlapping with s-wave resonance by the laser field.}. \textbf{(a)}
Loss measurements of the p-wave Feshbach resonance of
$|9/2,-7/2\rangle\otimes|9/2,-7/2\rangle$ as the function of the
magnetic field for the red laser detuning. \textbf{(b)} Loss
measurements of the p- and s-wave Feshbach resonances of
$|9/2,-7/2\rangle\otimes|9/2,-9/2\rangle$ as the function of the
magnetic field for the red laser detuning. The laser beam
propagating along the $\hat{y}$ axis, is linearly polarized parallel
to external magnetic field and red detuning with 1.1 GHz. }
\label{sp}
\end{figure}

\textit{Overlapping $s$- and $p$-wave Resonance.} Recently, several
works have predicted that interesting many-body physics can occur
when a $p$-wave Feshbach resonance sits nearby an $s$-wave one
\cite{cui16,cui-s-p,Yi2016,jiang2016}. For instance, it has been
predicted that for a one-dimensional Fermi gas with strong $s$-wave
interaction, an extra $p$-wave interaction can make the system favor
an itinerant ferromagnetic phase \cite{cui16, jiang2016}, providing
a new mechanism for itinerant ferromagnetism; and it has also been
predicted that interesting pairing structure can happen for a
three-dimensional Fermi gas with overlapping $s$- and $p$-wave
resonances. Nevertheless, without the optical control, even through
in ${}^{40}$K the $s$- and $p$-wave resonances are quite close, the
$p$-wave resonances sitting at $198.3$ G and $198.8$ G are barely
within the range ($8$ G) of the $s$-wave resonance sitting at
$201.6$ G \cite{s-wave-jin}.

With our optical control, as shown in Fig. \ref{sp}, for red
detuning $1.1$ GHz, one of the resonance with $m=0$ can be shifted
by about $10$ G in practice, and therefore overlaps with the
$s$-wave resonance. Fig \ref{sp} (a) shows the loss for single
component Fermi gas with only $|9/2,-7/2\rangle$ state and with
applied optical control, and Fig. \ref{sp}(b) shows the loss feature
for a mixture of $|9/2,-7/2\rangle$ and $|9/2,-9/2\rangle$. One can
see that one of the $p$-wave resonance is entirely buried inside the
$s$-wave resonance. Therefore it creates the situation where the
predications from Ref. \cite{cui16,cui-s-p,Yi2016,jiang2016} can be
tested in this system.

\textit{Conclusion.} In summary, we have studied the optical control of a $p$-wave Feshbach resonance by utilizing bound-to-bound
transitions driven by a laser field. The main finding is a universal
feature of this optical control, that is, the ratio $I(0)/I(\pm 1)$
to large extent is a universal constant. By this optical control, we
demonstrate that intriguing scenarios can happen such as a
$p$-wave resonance can overlap with an $s$-wave resonance. We have also considered the situation that
the polarization of the laser is not along $\hat{z}$ but in the $xy$ plane. This breaks the rotational symmetry and all
three resonances will split. This allows us to access independent control of all three resonances \cite{sup}. Our work
opens many opportunities for investigating interesting few- and
many-body problems in these settings.

\textit{Acknowlegment.} This work is supported by MOST under Grant
No. 2016YFA0301600 and No. 2016YFA0301602, NSFC Grant No. 11234008,
No. 11474188, No. 11704234, No. 11325418, No. 11734010, No.
11434011, No. 11674393, the Fund for Shanxi "1331 Project" Key
Subjects Construction, the Fundamental Research Funds for the
Central Universities, and the Research Funds of Renmin University of
China under Grant No.16XNLQ03, 17XNH054.

\end{document}


\title{Supplementary for ``Universal Feature in Optical Control of a $p$-wave Feshbach Resonance''}

\maketitle

\section{experimental setup}
We perform our experiments by employing a fermionic gas of $^{40}$K
atoms of the $F=9/2$ manifold. The experiment starts with a Fermi
gas of $|9/2,9/2\rangle$ with atom number $N=2\times10^{6}$ and
at a temperature of $T/T_{F}\approx0.3$ in a crossed 1064 nm optical
dipole trap . $T_{F}$ is the Fermi temperature defined as $T_{F}=(6N)^{1/3}\hbar\overline{\omega}/k_{B}$
with $\bar{\omega}\simeq2\pi\times80$ Hz labels the geometric trapping
frequency. The fermionic atoms are transferred to $|9/2,-9/2\rangle$ state
as the initial state via a rapid adiabatic passage induced by a RF
field at $5$ G. Then, the Fermi gas is transferred to the $|9/2,-7/2\rangle$
state using a RF field with duration of 30 $ms$ at $B\simeq219.4$
$\emph{G}$, where the frequency of center is 47.45 $\emph{MHz}$
and the width is 0.3 MHz. In addition to these, we
can also prepare the Fermi gases at the $|9/2,-5/2\rangle$ state
via transferring atoms in $|9/2,-7/2\rangle$ state to the $|9/2,-5/2\rangle$
state by a RF field of $\pi$ pulse.

Subsequently, a homogeneous magnetic bias field $B_{exp}$ is applied
in the $\hat{z}$ axis (gravity direction) produced by quadrupole
coils, which is operating in the Helmholtz configuration. Two laser
beams propagating along $\hat{y}$ and $\hat{z}$ respectively are
used as the tools to manipulate the $p$-wave Feshbach resonance
and are extracted from a continuous-wave Ti-sapphire single frequency
laser and focused at the position of the atomic cloud with $1/e^{2}$
radii of 200 $\mu m$, larger than the size of the degenerate Fermi
gas. The laser beams is frequency-shifted by an acousto-optic modulators
(AOM), which allows precise control of the laser intensity and duration
time of the pulse.

In order to control and observe the $p$-wave Feshbach resonance,
we start with the ultracold Fermi gases in the $|9/2,-7/2\rangle$
state at $B\simeq219.4$ $\emph{G}$. Then we adiabatically ramp the
magnetic-field to the various expected field $B_{\rm  exp}$ during 1 ms,
and hold 20 ms to observe the atomic losses at $|9/2,-7/2\rangle$
state by counting the number of atoms. Subsequently, the laser is
switched on and couples the closed channel molecular state to the
excited molecular states as shown in Fig. 1(a). Finally, we immediately
turn off laser beam, the optical trap, and the magnetic field, and
let the atoms ballistically expand in 12 ms and take the time-of-flight
(TOF) absorption image. The number of atoms in $|9/2,-7/2\rangle$
state is obtained from the TOF image.

\begin{figure}[t]
\begin{centering}
\includegraphics[clip,width=5in]{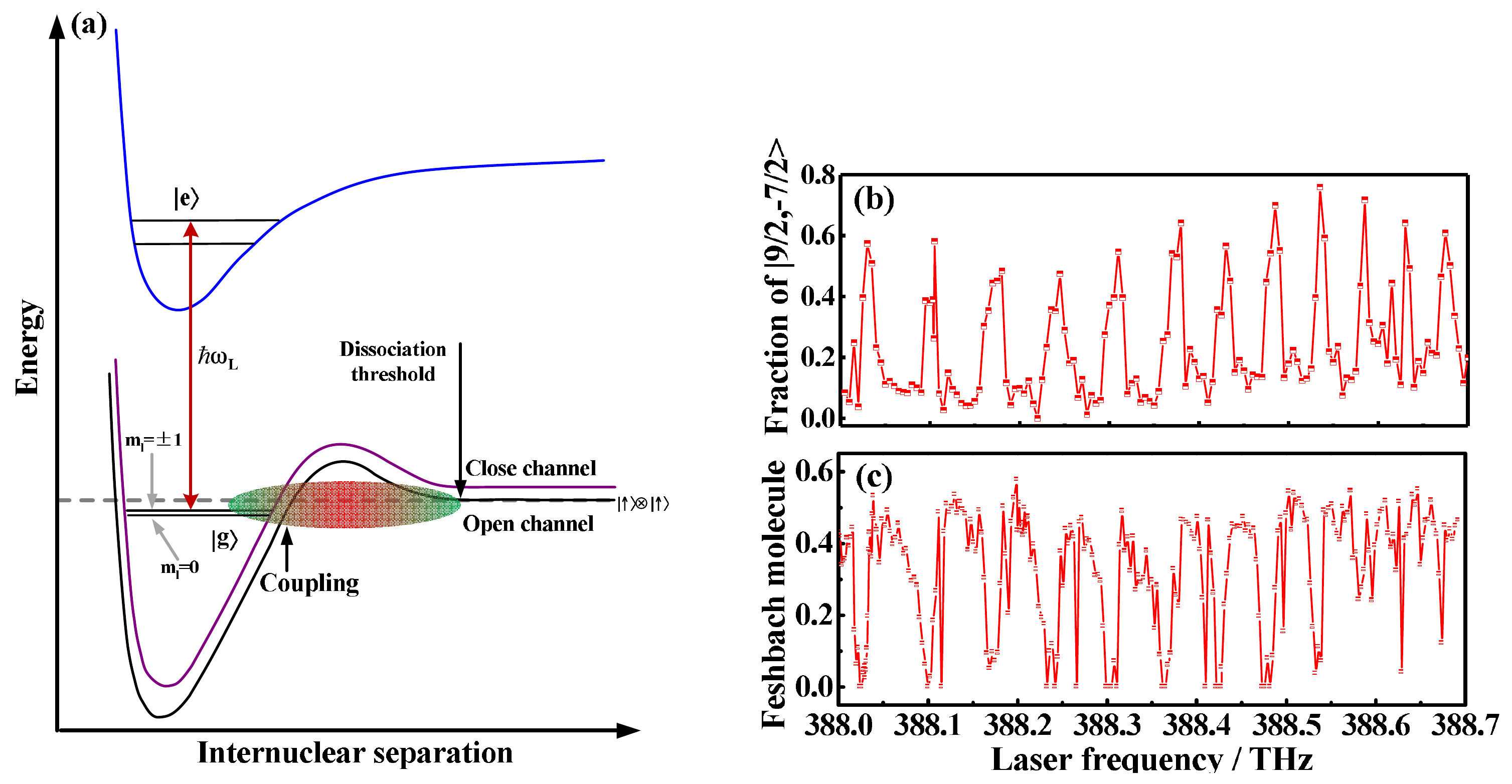}
\par\end{centering}
\caption{(Color online) \textbf{Energy level diagram and excited molecule  state
spectroscopy}. \textbf{(a)} Schematic diagram of the energy curves
of two atoms. Bound-to-bound transition from closed channel molecular
state $\phi_{{\rm g}}$ to one of the excited molecular states $\phi_{{\rm e}}$
occurs when a near resonant laser with frequency $\omega_{L}$ is
applied. Tow curves at the low energy end denotes the energy of two
single-component $^{40}$K atoms in electronic ground state . The
curve at high energy end denote the energy of two atoms composed by
one electronic ground state atom and one electronic excited state
atom.\textbf{(b)} Bound-to-bound spectroscopy below the $^{2}P_{1/2}+^{2}S_{1/2}$
threshold near $p$-wave Feshbach resonance of $|9/2,-7/2\rangle\otimes|9/2,-7/2\rangle$
at the magnetic field $B=198.3$ G. \textbf{(c)} Bound-to-bound spectroscopy
below the $^{2}P_{1/2}+^{2}S_{1/2}$ threshold near $s$-wave Feshbach
resonance of $|9/2,-9/2\rangle\otimes|9/2,-7/2\rangle$ at the magnetic
field $B=201.6$ G.}

\label{fig1}
\end{figure}

\section{bound-to-bound spectroscopy}

We first measure the bound-to-bound spectroscopy for excited
$^{40}K_{2}$ molecules below the $^{2}P_{1/2}+^{2}S_{1/2}$ threshold
near $p$-wave Feshbach resonance. The magnetic-field $B_{\rm exp}$
is set to 198.3 G. At this value the atoms are subject to inelastic
loss since the energy of the closed channel molecular state $m=\pm1$
coincides with the energy of two free atoms. When the laser
illuminates atomic gas and is near resonant with a bound-to-bound
transition from $\phi_{{\rm g}}$ to one of the excited molecular
states $\phi_{{\rm e}}$, a shift of the resonance position is
induced by ac-stark effect and the peak location of atomic losses
are shifted. Fig. \ref{fig1}(b) shows the bound-to-bound
spectroscopy near $p$-wave Feshbach resonance. Here, the laser
intensity is $I=60$ mW and the laser wavelength ranges from $771.5$
nm to $772.7$ nm. We compare this bound-to-bound spectroscopy near
$p$-wave Feshbach resonance with that near $s$-wave Feshbach
resonance of $|9/2,-9/2\rangle\otimes|9/2,-7/2\rangle$ at the
magnetic field $B=201.6$ G as shown in Fig. 1(c). This
bound-to-bound spectroscopy near $s$-wave Feshbach resonance was
reported in our earlier work \cite{Fu2013}, in which Feshbach
molecules are prepared below the resonance at $B=201.6$ G and its
losses are measured during the laser drive the bound-to-bound
transition. The bound-to-bound spectroscopy obtained by two
different ways show highly consistency. The observed peaks in the
bound-to-bound spectroscopy near $p$-wave Feshbach resonance
correspond to the vibrational level of the excited molecular states.
There should be the multi-substructures at each vibrational level
induced by vibration, rotation, hyperfine interaction, and Zeeman
interaction of molecules, which were observed in the bound-to-bound
spectroscopy near $s$-wave Feshbach resonance in our earlier work
\cite{Fu2013}.

%
%
%

\section{Proof of Eqs. (3) and (4)}
Now we prove Eq. (3) and Eq. (4) in the main text by calculate the intensity $I(m)$ of the laser-induced coupling between the closed-channel bound state and the excited molecular state.

\begin{figure}[t]
\begin{centering}
\includegraphics[clip,width=2.5in]{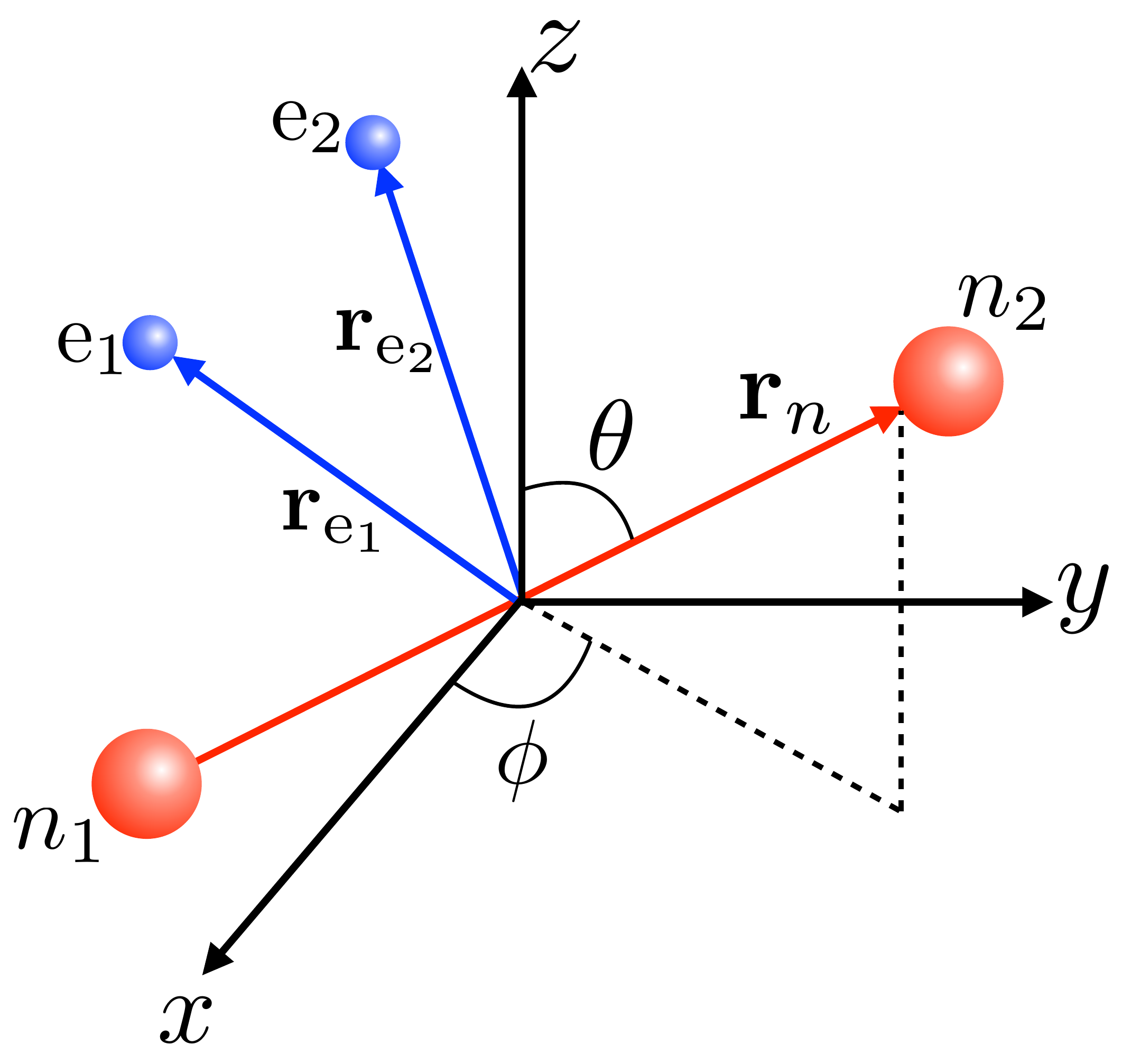}
\par\end{centering}
\caption{(Color online) The coordinate system used in our calculation.}
\label{fig2}
\end{figure}

We start with a  brief
introduction of the wave function of the molecular state $|l,m,\lambda,n_{\rm n},\{n_{\rm e}\}\rangle$. Since the freedom of the inner shell electron can be safely ignored,
a homonuclear diatomic molecule, {\it e.g.}, a molecule of two $^{40}$K atoms can be viewed as
being composed by two nuclei ${\rm n}_{1,2}$ and  two outermost shell electrons ${\rm e}_{1,2}$ (Fig.~\ref{fig2}).
We study the relative motion of these four particles, which is decoupled from the center-of-mass motion.
Thus, as shown in Fig.~\ref{fig2}, we choose the origin of our coordinate system to be the center-of-mass position, which is approximated as the
middle point of the two nuclei. We also define the $x$-, $y$- and $z$-axises to be parallel to the ones of the lab
frame.
In our system the
molecule wave function is a function of the relative position ${\bf r}_{\rm n}$ of the two nuclei and the
position ${\bf r}_{{\rm e}_{1,2}}$  of the
electrons ${\rm e}_{1,2}$(Fig.~\ref{fig2}). Notice that ${\bf r}_{{\rm e}_i}$($i=1,2$) is actually the relative position of ${\rm e}_{i}$
and the center of mass of the two nuclei. As shown in the main text, spin-orbit coupling and hyperfine interaction are ignored. Thus in our calculation we only consider the spatial motion of the electrons and nuclei.

In our system, the total orbital angular momentum  of all the four particles is denoted by  ${\bf L}$, and the orbital angular momentum of the electron ${\rm e}_i$ ($i=1,2$) is denoted by  ${\bf L}_{{\rm e}_i}$.
For simplicity, we ignore the fine and hyperfine interaction and only consider the Coulomb interaction. Therefore, for our system the total angular momentum ${\bf L}$ and the component of ${\bf L}_{{\rm e}_1}+{\bf L}_{{\rm e}_2}$ in the direction of ${\bf r}_{\rm n}$ are conserved.   Thus, the molecule state can be denoted as $|l,m,\lambda,n_{\rm n},\{n_{\rm e}\}\rangle$, where
$l$ and $m$ are the quantum numbers for ${\bf L}^2$ and the $z$-component of ${\bf L}$, respectively,
$\lambda$ is the quantum number for the component of
${\bf L}_{{\rm e}_1}+{\bf L}_{{\rm e}_2}$ along the direction of ${\bf r}_{\rm n}$,
while $n_{{\rm n}}$ and $\{n_{{\rm e}}\}$
are the nuclear and the electronic vibrational quantum numbers, respectively. Under the Born adiabatic approximation \cite{rs}, the molecular
wave function
\begin{equation}
\Psi_{l,m,\lambda,n_{\rm n},\{n_{\rm e}\}}({\bf r}_{{\rm n}};{\bf r}_{{\rm e}_{1}},{\bf r}_{{\rm e}_{2}})\equiv\langle
{\bf r}_{{\rm n}};{\bf r}_{{\rm e}_{1}},{\bf r}_{{\rm e}_{2}}|l,m,\lambda,n_{\rm n},\{n_{\rm e}\}\rangle
\end{equation}
can be factorized as
\begin{equation}
\Psi_{l,m,\lambda,n_{\rm n},\{n_{\rm e}\}}({\bf r}_{{\rm n}};{\bf r}_{{\rm e}_{1}},{\bf r}_{{\rm e}_{2}})=\psi_{l,m,\lambda,n_{\rm n},\{n_{\rm e}\}}^{({\rm n})}({\bf r}_{{\rm n}})\psi_{\{n_{{\rm e}}\},\lambda}^{({\rm e})}({\bf r}_{{\rm n}};{\bf r}_{{\rm e}_{1}},{\bf r}_{{\rm e}_{2}}).\label{mw1}
\end{equation}
Here $\psi_{\{n_{{\rm e}}\},\lambda}^{({\rm e})}$
is the wave function of the two outermost shell electrons when the
positions of the two nuclei are pinned down, and can be expressed as
\begin{equation}
\psi_{\{n_{{\rm e}}\},\lambda}^{({\rm e})}({\bf r}_{{\rm n}};{\bf r}_{{\rm e}_{1}},{\bf r}_{{\rm e}_{2}})=e^{-i\hat{L}_{z}^{({\rm e})}\phi}e^{-i\hat{L}_{y}^{({\rm e})}\theta}\phi_{\{n_{{\rm e}}\},\lambda}(r_{{\rm n}};{\bf r}_{{\rm e}_{1}},{\bf r}_{{\rm e}_{2}}),\label{ew}
\end{equation}
where $r_{{\rm n}}$, $\theta$ and $\phi$ are the norm, polar angle
and azimuthal angle of ${\bf r}_{{\rm n}}$, respectively (Fig.~\ref{fig2}), $\hat{L}_{\alpha}^{({\rm e})}$ ($\alpha=x,y,z$) is the component
of ${\bf L}_{{\rm e}_1}+{\bf L}_{{\rm e}_2}$ along
the $\alpha$-axis, and $\phi_{\{n_{{\rm e}}\},\lambda}$ is the electronic
wave function when the nuclei are pinned on the $z$-axis, which satisfies $\hat{L}_{z}^{({\rm e})}\phi_{\{n_{{\rm e}}\},\lambda}(r_{{\rm n}};{\bf r}_{{\rm e}_{1}},{\bf r}_{{\rm e}_{2}})=\lambda\phi_{\{n_{{\rm e}}\},\lambda}(r_{{\rm n}};{\bf r}_{{\rm e}_{1}},{\bf r}_{{\rm e}_{2}}).$
Moreover, it can be shown that the nuclear wave function $\psi_{l,m,\lambda,n_{\rm n},\{n_{\rm e}\}}({\bf r}_{{\rm n}})$
can be further factorized as \cite{landau}
\begin{equation}
\psi_{l,m,\lambda,n_{\rm n},\{n_{\rm e}\}}({\bf r}_{{\rm n}})=\chi_{l,m,\lambda,n_{\rm n},\{n_{\rm e}\}}(r_{{\rm n}})\sqrt{\frac{2l+1}{4\pi}}D_{\lambda,m}^{(l)}(\phi,\theta,0),\label{nw}
\end{equation}
with $D_{\lambda,m}^{(l)}(\alpha,\beta,\gamma)$ being the Wigner's
$D$-function and $\chi_{l,m,\lambda,n_{\rm n},\{n_{\rm e}\}}(r_{\rm n})$
being the radial wave function of the nuclei.

With the above results we can calculate the dipole transition matrix element $\langle l^\text{f}, m, \lambda^\text{f}, n_\text{n}^\text{f}, \{n^\text{f}_\text{e}\}|\hat{T}_{0}|l^\text{i}, m, \lambda^\text{i}, n_\text{n}^\text{i}, \{n^\text{i}_\text{e}\}\rangle$, which appears in Eq. (2) of the main text. This matrix element can be expressed as
\begin{eqnarray}
&&\langle l^\text{f}, m, \lambda^\text{f}, n_\text{n}^\text{f}, \{n^\text{f}_\text{e}\}|\hat{T}_{0}|l^\text{i}, m, \lambda^\text{i}, n_\text{n}^\text{i}, \{n^\text{i}_\text{e}\}\rangle\nonumber\\
&=&\int d{\bf r}_{{\rm e}_{1}}d{\bf r}_{{\rm e}_{2}}d{\bf r}_{{\rm n}}\left[\Psi_{l^\text{f}, m, \lambda^\text{f}, n_\text{n}^\text{f}, \{n^\text{f}_\text{e}\}}^{\ast}({\bf r}_{{\rm n}};{\bf r}_{{\rm e}_{1}},{\bf r}_{{\rm e}_{2}})\hat{T}_{0}\Psi_{l^\text{i}, m, \lambda^\text{i}, n_\text{n}^\text{i}, \{n^\text{i}_\text{e}\}}({\bf r}_{{\rm n}};{\bf r}_{{\rm e}_{1}},{\bf r}_{{\rm e}_{2}})\right],\label{omega}
\end{eqnarray}
where
\begin{equation}
\hat{T}_{0}=(z_{{\rm e}_{1}}+z_{{\rm e}_{2}}),\label{t0}
\end{equation}
with $z_{{\rm e}_{i}}$
($i=1,2$) is the $z$-component of ${\bf r}_{{\rm e}_{i}}$. For future use, we further
define the operators
\begin{equation}
\hat{T}_{\pm1}=\left[\frac{\mp(x_{{\rm e}_{1}}+x_{{\rm e}_{2}})-i(y_{{\rm e}_{1}}+y_{{\rm e}_{2}})}{\sqrt{2}}\right],\label{tpm1}
\end{equation}
with $x_{{\rm e}_{i}}$ and $y_{{\rm e}_{i}}$ ($i=1,2$) being the
$x$- and $y$-component of ${\bf r}_{{\rm e}_{i}}$, respectively.
It is clear that $\hat{T}_{j}$ ($j=0,\pm1$) form rank-1 irreducible
tensor operators under the total rotation of the two electrons, and
thus satisfy
\begin{equation}
e^{i\hat{L}_{y}^{({\rm e})}\theta}e^{i\hat{L}_{z}^{({\rm e})}\phi}\hat{T}_{0}e^{-i\hat{L}_{z}^{({\rm e})}\phi}e^{-i\hat{L}_{y}^{({\rm e})}\theta}=\sum_{q'=0,\pm1}\hat{T}_{q'}D_{q',0}^{(1)}(\phi,\theta,0).\label{irt}
\end{equation}
Substituting Eqs.(\ref{ew},\ref{nw}) into Eq.(\ref{mw1})
and then into Eq.(\ref{omega}), and using Eq. (\ref{irt}) and the facts that $l^{\rm i}=1$ and $\lambda^{\rm i}=0$, we obtain
\begin{equation}
\langle l^\text{f}, m, \lambda^\text{f}, n_\text{n}^\text{f}, \{n^\text{f}_\text{e}\}|\hat{T}_{0}|l^\text{i}, m, \lambda^\text{i}, n_\text{n}^\text{i}, \{n^\text{i}_\text{e}\}\rangle=(-1)^{\lambda^{\rm f}-m}\sqrt{3(2l^{\rm f}+1)}
A_{\{n_{\rm e}^{\rm f}\},n_{{\rm n}}^{\rm f},\{n_{\rm e}^{\rm i}\},n_{{\rm n}}^{\rm i},l^{\rm f},m,\lambda^{\rm f}}
\left(\begin{array}{ccc}
l^{\rm f} & 1 & 1\\
-\lambda^{\rm f} & \lambda^{\rm f} & 0
\end{array}\right)\left(\begin{array}{ccc}
l^{\rm f} & 1 & 1\\
-m & 0 & m
\end{array}\right),\label{omega3}
\end{equation}
where $\left(\begin{array}{ccc}
j_{1} & j_{2} & j_{3}\\
m_{1} & m_{2} & m_{3}
\end{array}\right)$ is the Winger-3$j$ symbols and $A_{\{n_{\rm e}^{\rm f}\},n_{{\rm n}}^{\rm f},l^{\rm f},m,\lambda^{\rm f}}$ is defined as
\begin{align}
& A_{\{n_{\rm e}^{\rm f}\},n_{{\rm n}}^{\rm f},\{n_{\rm e}^{\rm i}\},n_{{\rm n}}^{\rm i},l^{\rm f},m,\lambda^{\rm f}}=\nonumber\\
 &\int_{0}^{\infty}dr_{{\rm n}}\left\{ r_{{\rm n}}^{2}\chi_{l^{\rm f},m,\lambda^{\rm f},\{n_{\rm e}^{\rm f}\},n_{{\rm n}}^{\rm f}}^{\ast}(r_{{\rm n}})\chi_{l^{\rm i}=1,m,\lambda^{\rm i}=0,\{n_{\rm e}^{\rm i}\},n_{{\rm n}}^{\rm i}}(r_{\rm n})\int d{\bf r}_{{\rm e}_{1}}d{\bf r}_{{\rm e}_{2}}\left[\phi_{\{n_{\rm e}^{\rm f}\},\lambda^{\rm f}}^{\ast}(r_{{\rm n}};{\bf r}_{{\rm e}_{1}},{\bf r}_{{\rm e}_{2}})\hat{T}_{\lambda^{\rm f}}\phi_{\{n_{\rm e}^{\rm i}\},\lambda^{\rm i}=0}(r_{{\rm n}};{\bf r}_{{\rm e}_{1}},{\bf r}_{{\rm e}_{2}})\right]\right\}.\nonumber\\
 \label{biga}
\end{align}
In the derivation of Eq. (\ref{omega3}), the relations that $D_{m'm}^{(l)}(\phi,\theta,0)^{*}=(-1)^{m'-m}D_{-m',-m}^{(l)}(\phi,\theta,0)$
and
\begin{equation}
\int\frac{\sin\phi d\phi d\theta}{4\pi}D_{m_{1}'m_{1}}^{(j_{1})}(\phi,\theta,0)D_{m_{2}'m_{2}}^{(j_{2})}(\phi,\theta,0)D_{m_{3}'m_{3}}^{(j_{3})}(\phi,\theta,0)=\left(\begin{array}{ccc}
j_{1} & j_{2} & j_{3}\\
m_{1}' & m_{2}' & m_{3}'
\end{array}\right)\left(\begin{array}{ccc}
j_{1} & j_{2} & j_{3}\\
m_{1} & m_{2} & m_{3}
\end{array}\right).\label{dr}
\end{equation}
are adopted\cite{landau}.

Now we consider the dependence of $A_{\{n_{\rm e}^{\rm f}\},n_{{\rm n}}^{\rm f},\{n_{\rm e}^{\rm i}\},n_{{\rm n}}^{\rm i},l^{\rm f},m,\lambda^{\rm f}}$ on the quantum numbers $l^{\rm f}$ and $m$ of the finial state. In the radial Schr$\ddot{{\rm o}}$dinger equation satisfied
by $\chi_{\{n_{\rm e}^{\rm f}\},n_{{\rm n}}^{\rm f},l^{\rm f},m,\lambda^{\rm f}}(r_{{\rm n}})$, the value
of $l^{\rm f}$ and $m$ only influence the intensity of the centrifugal
potential which is proportional to $-r_{{\rm n}}^{-2}$, and thus
do not have strong effects for the deep bound state. As shown in the main text, here we assume
that the excited molecular states are deep bound states and thus assume
the radial wave function $\chi_{\{n_{\rm e}^{\rm f}\},n_{{\rm n}}^{\rm f},l^{\rm f},m,\lambda^{\rm f}}(r_{{\rm n}})$
to be approximately independent of the values of $l'$ and $m$.
Therefore, according to Eq. (\ref{biga}), the parameter $A_{\{n_{\rm e}^{\rm f}\},n_{{\rm n}}^{\rm f},\{n_{\rm e}^{\rm i}\},n_{{\rm n}}^{\rm i},l^{\rm f},m,\lambda^{\rm f}}$
is also independent of $l^{\rm f}$ and $m$. Therefore, we can simplify the notation
\begin{eqnarray}
A_{\{n_{\rm e}^{\rm f}\},n_{{\rm n}}^{\rm f},\{n_{\rm e}^{\rm i}\},n_{{\rm n}}^{\rm i},l^{\rm f},m,\lambda^{\rm f}}\rightarrow A_{\{n_{\rm e}^{\rm f}\},n_{{\rm n}}^{\rm f},\{n_{\rm e}^{\rm i}\},n_{{\rm n}}^{\rm i},\lambda^{\rm f}},
\end{eqnarray}
and rewrite Eq. (\ref{omega3}) in a more concise form as
\begin{equation}
\langle l^\text{f}, m, \lambda^\text{f}, n_\text{n}^\text{f}, \{n^\text{f}_\text{e}\}|\hat{T}_{0}|l^\text{i}, m, \lambda^\text{i}, n_\text{n}^\text{i}, \{n^\text{i}_\text{e}\}\rangle=(-1)^{\lambda^{\rm f}-m}\sqrt{3(2l^{\rm f}+1)}
A_{\{n_{\rm e}^{\rm f}\},n_{{\rm n}}^{\rm f},\{n_{\rm e}^{\rm i}\},n_{{\rm n}}^{\rm i},\lambda^{\rm f}}
\left(\begin{array}{ccc}
l^{\rm f} & 1 & 1\\
-\lambda^{\rm f} & \lambda^{\rm f} & 0
\end{array}\right)\left(\begin{array}{ccc}
l^{\rm f} & 1 & 1\\
-m & 0 & m
\end{array}\right).\label{omega4}
\end{equation}
Eq. (\ref{omega4}) shows
that the dependence of $\langle l^\text{f}, m, \lambda^\text{f}, \{n^\text{f}_\text{e}\}, n_\text{n}^\text{f}|\hat{T}_{0}|l^\text{i}, m, \lambda^\text{i}, \{n^\text{i}_\text{e}\}, n_\text{n}^\text{i}\rangle$
on $l^{\rm f}$ and $m$ are all included in the Winger-3$j$
symbols and the factor $\sqrt{3(2l^{\rm f}+1)}$, and thus can be evaluated precisely.

Substituting Eq. (\ref{omega4}) into Eq. (2) of the main text, we immediately obtain the laser coupling intensity $I(m)$:
\begin{eqnarray}
\label{Im2}
I(m)=g(m,\lambda^\text{f})\times h\left(\{n^\text{f}_\text{e}\},n_\text{n}^\text{f},\{n^\text{i}_\text{e}\},n_\text{n}^\text{i},\lambda^{\rm f}\right),\label{emt}
\end{eqnarray}
where
\begin{eqnarray}
g(m,\lambda^{\rm f})=\sum_{l^{\rm f}}3(2l^{\rm f}+1)
\left(\begin{array}{ccc}
l^{\rm f} & 1 & 1\\
-\lambda^{\rm f} & \lambda^{\rm f} & 0
\end{array}\right)^{2}\left(\begin{array}{ccc}
l^{\rm f} & 1 & 1\\
-m & 0 & m
\end{array}\right)^{2},
\end{eqnarray}\label{15}
 and
 \begin{eqnarray}
 h\left(\{n^\text{f}_\text{e}\},n_\text{n}^\text{f},\{n^\text{i}_\text{e}\},n_\text{n}^\text{i},\lambda^{\rm f}\right)=\left|A_{\{n_{\rm e}^{\rm f}\},n_{{\rm n}}^{\rm f},\{n_{\rm e}^{\rm i}\},n_{{\rm n}}^{\rm i},\lambda^{\rm f}}\right|^2.\label{16}
 \end{eqnarray}
Eq. (\ref{emt}) is just Eq.(3) of our main text.

Furthermore, using Eqs. (\ref{Im2}-\ref{16}) we can immediately obtain the result $I(+1)=I(-1)$, as well as Eq. (4) of our main text:
\begin{eqnarray}
\frac{I(0)}{I(\pm 1)}=\left\{ \begin{array}{c}
3,{\rm \ for\ \lambda^{\rm f}=0}\\
\\
1/2,{\rm \ for\ \lambda^{\rm f}=\pm1}
\end{array}\right..
\end{eqnarray}
Notice that this result is independent of the value of $\{n^{\rm f}_{\rm e}\}$. This result means that if the $\pi$-polarized
laser couples the closed-channel bound state of the magnetic Feshbach resonance to the excited molecular states with $\lambda^{\rm f}=0$ (i.e., the states in the electronic $\Sigma$-orbit), then the shift of the resonance point for $m=0$ is more significant than that for $m=\pm 1$. On the other hand, if the excited molecular states have $\lambda^{\rm f}=\pm 1$ (i.e., the states are in the electronic $\Pi$-orbit), then the shift of the resonance point for $m=\pm 1$ is more significant than that for $m=0$. This result implies in all of our experiments the laser induce the coupling to the excited molecule states with $\lambda^{\rm f}=0$.

\begin{figure}[t]
\begin{centering}
\includegraphics[clip,width=0.45\textwidth]{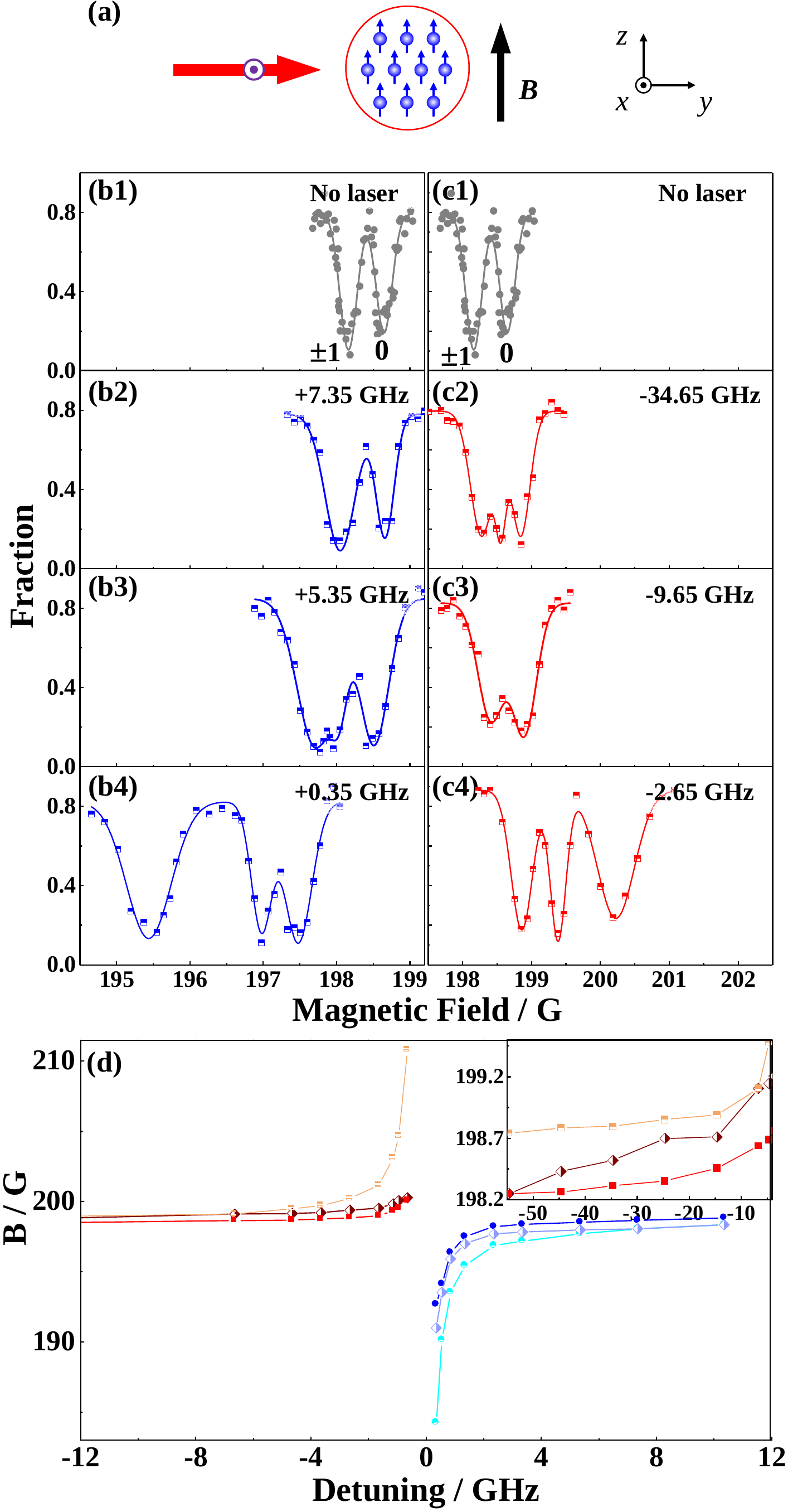}
\par\end{centering}
\caption{(Color online) \textbf{The $p$-wave Feshbach resonance
manipulated by the laser field with the polarization perpendicular
to external magnetic field.} \textbf{(a)} Schematic diagram of the
laser beam and the external magnetic field. The laser beam
propagating along the $\hat{y}$ axis, is linearly polarized
perpendicular to external magnetic field. \textbf{(b)} and
\textbf{(c)} Atom loss measurements of the p-wave Feshbach resonance
of $|9/2,-7/2\rangle\otimes|9/2,-7/2\rangle$ as the function of the
magnetic field for the different blue and red laser detuning.
\textbf{(d)} The resonance position of the shifted Feshbach
resonance as a function of the laser detuning. Three different lines
correspond to the $m=0$ and $m=\pm 1$ resonances respectively. }
\label{nosymmetry}
\end{figure}

\begin{figure}[t]
\begin{centering}
\includegraphics[width=0.45\textwidth]{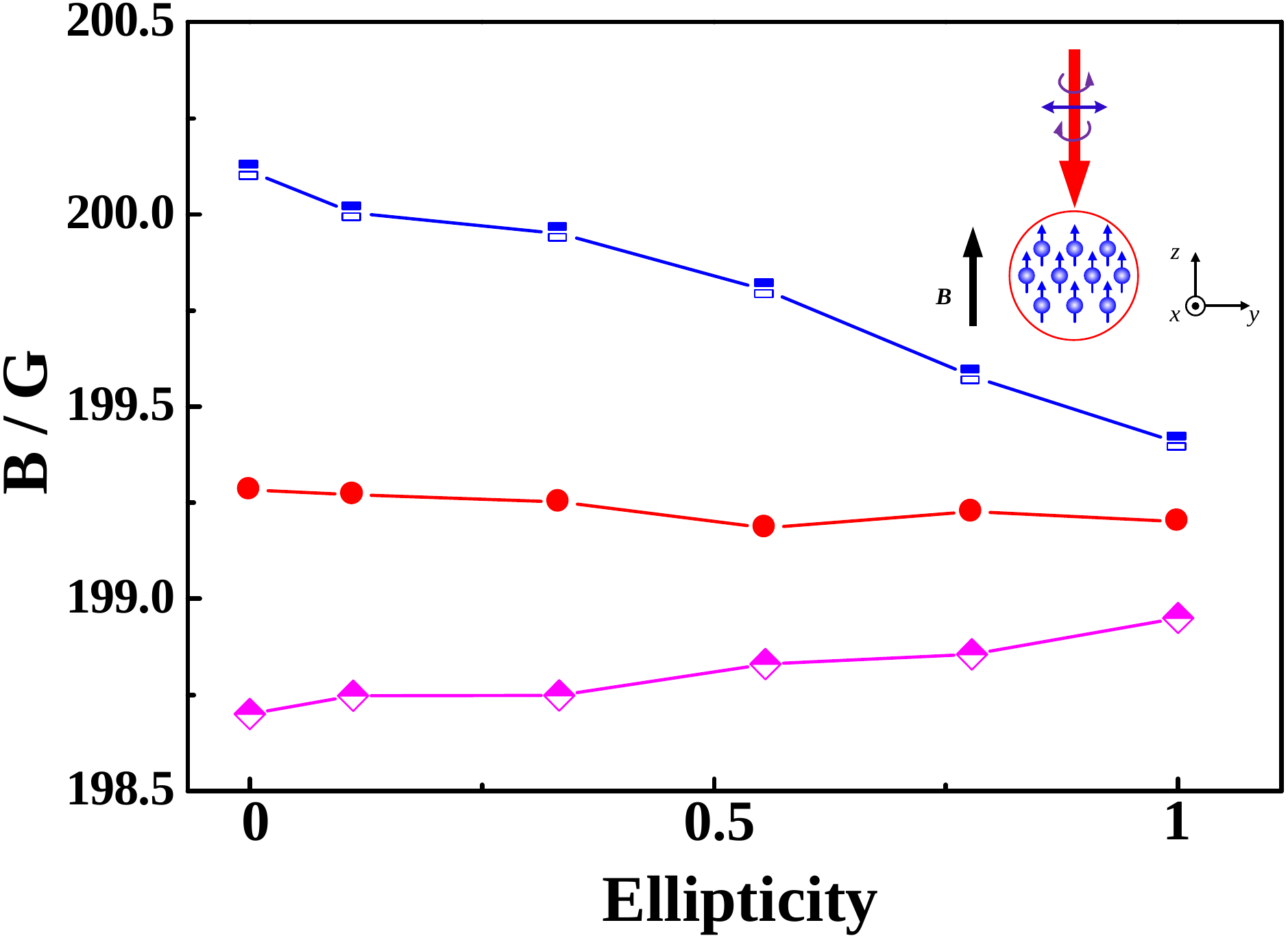}
\par\end{centering}

\caption{(Color online) \textbf{The p-wave Feshbach resonance
manipulated by the laser field propagating along the external
magnetic field with circular polarization.} The  position of the
shifted Feshbach resonance as a function of the ellipticity. Here,
laser detuning is $-2.6$ GHz. Inset: Schematic diagram of the laser
beam and the external magnetic field. Three different lines
correspond to the $m=0$ and $m=\pm 1$ resonances respectively.}

\label{ellipcity}
\end{figure}

\section{Varying Laser Polarizations.} 
We also consider the situation where the laser polarization is changed. In Fig.
~\ref{nosymmetry} we show the case where the laser propagates along
$\hat{y}$ and the polarization is along $\hat{x}$. In this case the
laser breaks rotational symmetry along $\hat{z}$ and the $m=\pm 1$
resonances split as a result.  We find that when the laser is red
detuned and large in strength, one of the peaks from $m=\pm 1$
manifold moves closer to $m=0$ resonance and they become
incidentally degenerate, as shown in Fig. \ref{nosymmetry}(c3). When
the laser frequency is tuned further to the resonance, the $m=0$
peak moves much more quickly, still consistent with the
analysis in main text.

In Fig. \ref{ellipcity} we show a different case in which the laser
propagates along the magnetic field direction but the polarization
is circularly polarized in the plane perpendicular to the magnetic
field. Here we fix the laser detuning at $-2.6$ GHz and find that
the resonance position behaves differently depending on the laser
ellipticity $\xi$. ($\xi=0$ denote linear polarization, and
$\xi=\pm$ denotes left and right circular polarization.) Thus, we
see that by combining laser detuning and ellipticity, one can almost
independently control all three resonances.